\documentclass[11pt] {article}
\usepackage{graphicx}
\usepackage{fancyhdr}
\usepackage{multicol}
\usepackage{styl-ap_mod}
\usepackage {epstopdf}
\begin{document}

\title {Oxygen rich cool stars in the Cepheus region, \\
New observations. III. }
\author{G.V. Petrosyan\work{1}, C. Rossi\work{2}, S. Gaudenzi\work{2.} R. Nesci\work{3}        }
\workplace{Yerevan State University, Armenia,
\next 
Dipartimento di Fisica, Universit\`a di Roma ``La Sapienza'', P.le Aldo Moro 5, 00185, Roma, Italy 
\next NAF/IAPS, via Fosso del Cavaliere 100, 00133 Roma, Italy}
\mainauthor  { e-mail: gohar.petrosyan@ysu.am }
\maketitle

\begin{abstract} We present moderate resolution CCD spectra and R photometry for seven KP2001 stars. We revised the spectral classification of the stars in the range $\lambda$~ 3900 - 8500~ \AA. On the bases of light curves of the NSVS (Northern Sky Variability Survey) database we classify KP2001-18 as a semi regular and KP2001-176 as Mira type variables. For all observed objects NSVS phase - dependence light curve analysis and variability type classification was performed with the VStar Software. Using period Ð luminosity relation we computed MK  magnitudes and the distances to variables.  
\end{abstract}

{\bf Key words:} stars: spectral classification: variables: distances  

\bigskip
\section {Introduction.} 
In this paper, third in this series, we continue spectroscopic and photometric study of oxygen-rich stars from KP2001 catalogue [1] in Cepheus direction. First two papers [2, 3] of the present series are devoted to clarify the nature of 20 late Ð type stars. In this paper we present moderate Ð resolution CCD spectra for seven stars from KP2001 catalogue. The luminosity classes and the distances are derived for them.  

 \section {Spectroscopic and photometric observations.} 
Moderate resolution CCD spectra for seven KP2001 stars are obtained in the range $\lambda$ 3900 - 8500~ \AA, dispersion 3.9/pixels, with the 1.52 m Cassini telescope of the Bologna Astronomical Observatory (Italy) at Loiano, equipped with the Bologna Faint Objects Spectrometer and Camera (BFOSC) and 1300x1300 pixel EEV P1129915 CCD. Photometric observations in R-band were also obtained with BFOSC in the same dates as the spectra. 
Table 1 presents the Journal of our observations, as well as the derived magnitudes and spectral types. The columns have the following meaning: column 1 Ð KP2001 number in the list [1], column 2 Ð date of observation, column 3 Ð R band magnitudes, column 4 Ð spectral types presented in KP2001 catalogue, column 5 - new spectral subtypes, revised from our CCD data, derived as described below. All the spectroscopic and photometric observations were processed by means of standard IRAF procedures.

\begin{table} 
\centering
\caption{JOURNAL OF OBSERVATIONS, PHOTOMETRIC RESULTS AND SPECTRAL CLASSIFICATION}
\begin{tabular}{ccccc}
\hline
\noalign{\smallskip}
KP2001 & date of & R mag & Previous Sp. & CCD Sp. \\
Number & observation &    &Subtype[1] & Subtype \\
\hline
  8   & 12.12.2014 & 13.5 $\pm$ 0.1 & M3 & M3III \\
  16 & 27.01.2015 & 13.2 $\pm$ 0.1 & M2 & M2III \\
  18 & 27.01.2015 & 10.8 $\pm$ 0.1 & M5 & M7III \\
176 & 12.12.2014 & 14.6 $\pm$ 0.1 & M6 & M8III \\
194 & 12.12.2014 & 11.9 $\pm$ 0.1 &M6 & M6III \\
243 & 29.12.2013 &  -                  & M4 & M2III \\
251 & 07.09.2011 &        -           & M7 & M10III \\
\hline
\end{tabular}
\end{table}

We determined spectral subtypes of stars via side-by side comparison with spectra of standard stars (from M0 to M9, giants and dwarfs) obtained with the same instrumentation.    Fig.1, 2 and 3 presents our CCD spectra for the seven KP2001 stars.

\begin{figure}
\centering
\includegraphics[width=10cm,height=10cm,angle=0]{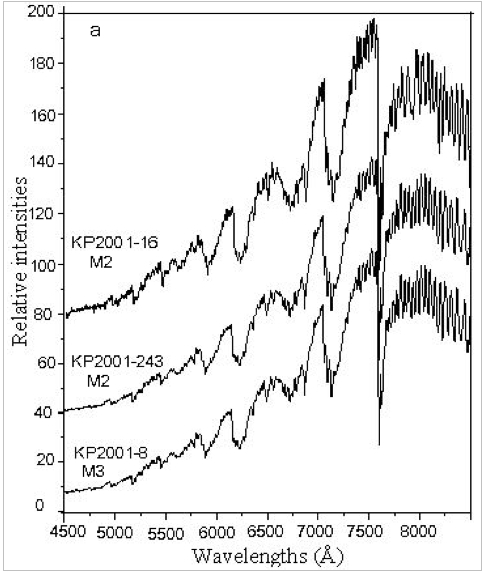}
\caption{Spectra of some KP2001 stars. Flux in arbitrary units, wavelengths in Angstrom.}
\end{figure}

\begin{figure}
\centering
\includegraphics[width=10cm,height=10cm,angle=0]{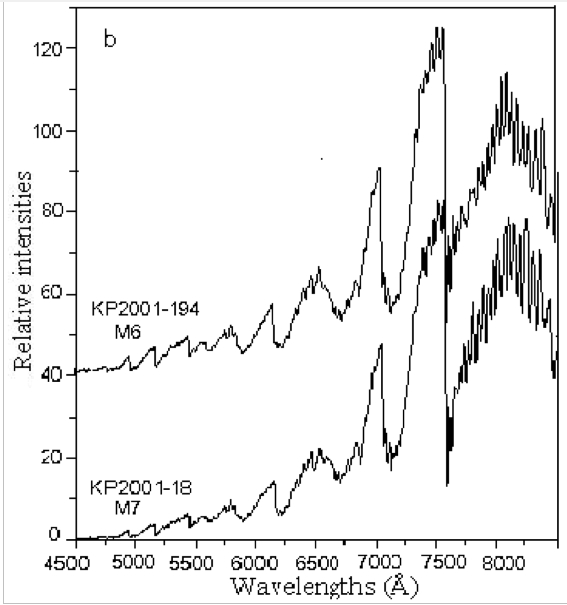}
\caption{Spectra of some KP2001 stars. Flux in arbitrary units, wavelengths in Angstrom.}
\end{figure}

\begin{figure}
\centering
\includegraphics[width=10cm,height=10cm,angle=0]{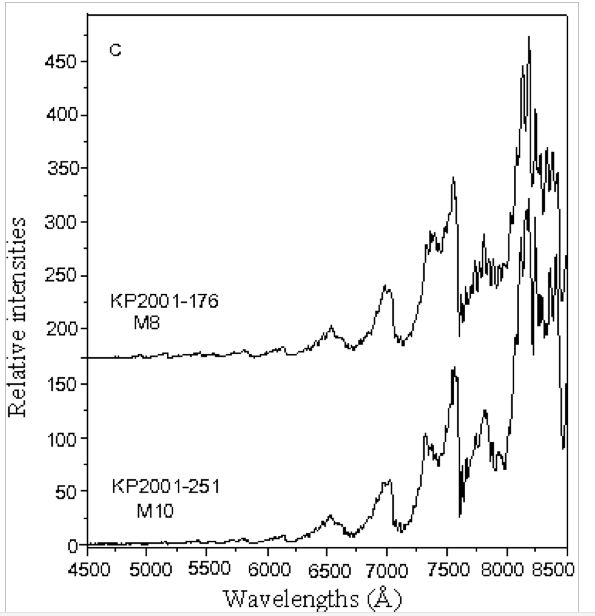}
\caption{Spectra of some KP2001 stars. Flux in arbitrary units, wavelengths in Angstrom.}
\end{figure}

\section{Colors, variability and distances.}

 We used the determined spectral classification to measure a number of physical properties of our stars.
\subsection{Colors.} 
The observed targets are located at low galactic latitudes, therefore the observed colors are strongly affected by interstellar reddening which is the first important quantity to be determined.  For this purpose we used the infrared colors, applying the same diagrams as in the papers [4-8]. A detailed discussion of the intrinsic colors of late type M Ð giants and dwarfs can be found in the paper by Bessell \& Brett [4]. 2MASS infrared (JHKS) magnitudes are available at http://irsa.ipac.caltech.edu/2mass. 
	In table 2 we present the original JHKS magnitudes from 2MASS catalogue for our stars.
	
\begin{table*} 
\centering
\caption{MAGNITUDES OF THE OBSERVED STARS}
\begin{tabular}{cccc}
\hline
\noalign{\smallskip}
KP2001  & J                 &   H                      &   K \\
Number   & mag.         & mag.                   & mag. \\
8      &9.056 $\pm$ 0.018  & 7.829 $\pm$ 0.031& 7.444 $\pm$ 0.018 \\ 
16   &          -                     & 7.398 $\pm$ 0.031 & 6.934 $\pm$ 0.027 \\
18   & 8.550 $\pm$ 0.020 & 3.923 $\pm$ 0.184 & 3.226 $\pm$ 0.242 \\
176 & 5.245 $\pm$ 0.238 & 6.770 $\pm$ 0.031 & 6.116 $\pm$ 0.022 \\
194 & 7.849 $\pm$ 0.035 & 5.643 $\pm$ 0.038 & 5.193 $\pm$ 0.024 \\
243 & 6.791 $\pm$ 0.021 & 5.933 $\pm$ 0.051 & 5.510 $\pm$ 0.021 \\
251 & 7.175 $\pm$ 0.025 & 3.125 $\pm$ 0.218 & 2.514 $\pm$ 0.304 \\
       & 4.296 $\pm$ 0.256 &    -                           &                        -   \\
\hline
\end{tabular}
\end{table*}

In Fig. 4 we reproduced the Fig. A3 of paper [4], where we added our stars after having transformed the original 2MASS magnitudes to the Bessell \& Brett system using the formulae given in Explanatory Supplement to the 2MASS Second Incremental Data Release at 

http://www.ipac.caltech.edu/2mass/release/second/doc/sec6-3.html and in Appendix A by Carpenter [9]. Fig. A3 of paper [4] is the (J-H) versus (H-K) diagram showing schematically the regions occupied by G5 to M6 dwarf and giants, SR and LPV carbon stars, and SR and LPV M7-M10 AGB stars. The arrow indicates the direction of interstellar reddening. 

\begin{figure}
\centering
\includegraphics[width=10cm,height=10cm,angle=0]{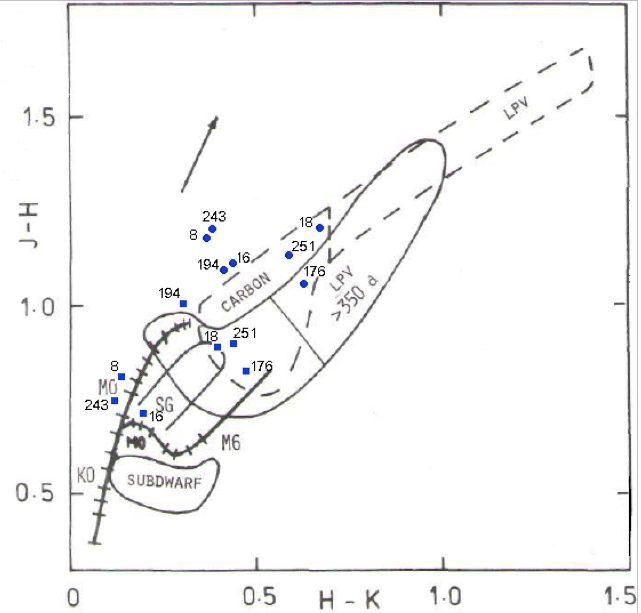}
\caption{J-H,H-K color-color diagram for our stars, labelled with their numbers}
\end{figure}

\begin{table} 
\centering
\caption{INFRARED COLORS AND COLOR EXCESS OF SEVEN KP2001 TARGETS TRANSFORMED TO THE BESSEL \& BRETT SYSTEM}
\begin{tabular}{cccccccc}
\hline
\noalign{\smallskip}
KP2001   & J-H  & E(J-H) & H-Ks & E(H-Ks) & E(B-V) & AV   & AK \\
number   & mag. & mag. & mag. & mag.      & mag.   & mag.&mag.\\
  8         & 1.29 $\pm$ 0.04 & 0.39 & 0.36 $\pm$ 0.04 & 0.12& 0.85&2.65&0.29\\
  16       &1.22$\pm$ 0.04&0.35&0.44$\pm$ 0.04&0.22&1.06&3.31&0.36\\
  18      &1.39$\pm$ 0.30&0.43&0.67$\pm$ 0.30&0.36&1.52&4.74&0.52\\
  176    &1.15$\pm$ 0.05&0.17&0.63$\pm$ 0.04&0.20&0.5-1.0&1.56+&0.15+\\
  194     &1.22$\pm$ 0.04&0.26&0.42$\pm$ 0.05&0.12&0.67&3.12&0.36\\
  243    &1.31$\pm$  0.06&0.44&0.39$\pm$ 0.06& 0.17 &0.81&2.09&0.23\\
  251   &1.24$\pm$0.34&0.22&0.58$\pm$0.37&0.10&0.56&2.53&0.27\\
           &                   &        &                    &       &      &1.75&0.19\\
\hline
\end{tabular}
\end{table}

We present in Table 3 the observed J-H and H-K$_s$ colors. From the spectral subtype determinations we deduced the expected intrinsic (reddened) values presented in Table III of paper [4] in order to move the experimental values to intrinsic positions and compute the extinction of the stars. For the Mira type stars KP2001-176 and 251 we used the relations between IR colors and logP given by [16] (formulae 1 and 2).
The observational position of KP2001-176 does not allow an accurate evaluation of the color excess. We therefore provide here the minimum and maximum values of the expected colors and of the derived quantities. Having obtained infrared-color excess, we computed E(B-V), A$_V$, and A$_K$, using the reddening relations given in the Appendix B of paper [4]; 

\medskip

     $ E(J-H) = 0.37 \cdot E(B-V);      E(H-K_s) = 0.19 \cdot E(B-V) $  ~~~~~~~~~~~~~~ (1)   

\medskip

$         A_V = 3.12 \cdot E(B-V);          A_K=0.34 \cdot E(B-V)       $   ~~~~~~~~~~~~~~~~~~~~~~~~~~~~~~~~~~~(2)    

\medskip

       In Fig. 4 infrared color Ð color diagrams are presented. Dots indicate the observed positions, squares indicate the positions after reddening correction. Some stars, investigated in paper [2] are presented in fig. 4 again. The color excess is presented in Table 2.


\subsection{Variability.}
 We considered phase dependence light curves of observed KP2001 stars from Northern Sky Variability Survey Ð NSVS [10] (online available at http:// skydot.lanl.gov/nsvs/ nsvs.php/) database to investigate their variability nature. For all observed objects NSVS phase Ð dependence light curve analysis and variability type classification was performed with the help of VStar Software (a multi latform data visualization and analysis tool, available at http://www.aavso.org/). VStar implements the Data Compensated Discrete Fourier Transform (DCDFT) algoritpl [11] to get the basic pulsation period. In the catalogue ÒRed variables in NSVSÓ [12] KP2001-251 (V0854 Cas) is classified as a Mira-type variable with period 347 days. This object is known also as OH maser source [13]. We classify KP2001-8 and 194 as Irregular (Irr) variables, NSVS light curves for KP2001-16 and 243 show no variability, we classify KP2001-18 as a Semi-Regular (SR, with period P = 60 days), and KP2001-176 and 251 as Mira-type variables with periods P= 320 days and P=347 days consequently.
In Fig. 5 and 6 presents NSVS phase dependence light curves for KP2001-18 and KP2001-176 consequently. The identification numbers of NSVS for KP2001-18 and 176 are respectively ID174401 and ID 187869.

\begin{figure}
\centering
\includegraphics[width=16cm,height=8cm,angle=0]{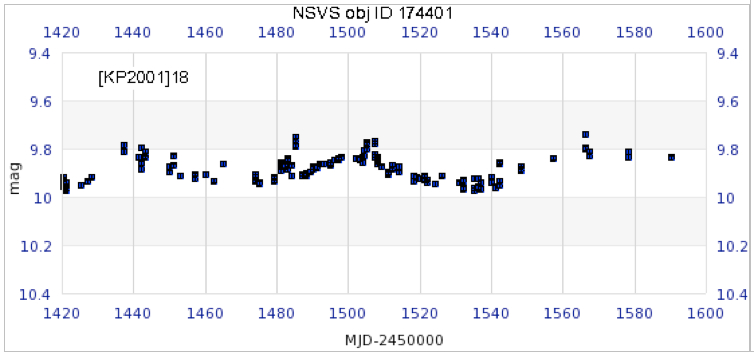}
\caption{NSVS phase dependence light curves for KP2001-18.}
\end{figure}

\begin{figure}
\centering
\includegraphics[width=16cm,height=8cm,angle=0]{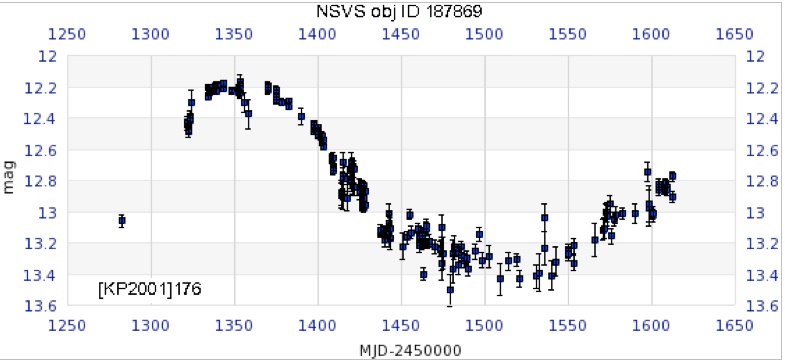}
\caption{NSVS phase dependence light curves for KP2001-176 star.}
\end{figure}

\subsection{Absolute magnitudes and Distances.}
To estimate the absolute magnitudes and distances to the variable stars we applied the same procedures of paper [1], using updated Period Ð Luminosity (PL) relations.  We first computed the absolute K magnitude for O Ð rich Mira variables applying the formula by [17]:

\medskip
    $M( K ) = -3.51 \pm 0.20 \cdot  (Log (P) - 2.38) -7.25 \pm 0.06  $ ~~~~~~~~~~~~~~~~~~~        (3)        

\medskip

 while for the   SR variable KP2001-18  we applied the  new PL relation given by [14]:

\medskip

$             M(K) = -1.34 \pm0.06  \cdot Log (P) - 4.5\pm 0.35  $ ~~~~~~~~~~~~~~~~~~~~~~~~~~~~~~ (4)   

\medskip
Then we have computed the distances, also taking into account the two possible values of the extinction in K magnitude for KP2001-176. For Mira Ð type variables the distances are estimated also using equation (3) presented in first Paper[ 2];

\medskip
              $  M(bol) = 2.80 -3.0 \cdot Log (P)  $ ~~~~~~  ~~~~~~~~ ~~~~~~~ ~~~~~~~  ~~~~~ ~~~~~~         (5)       
                                                                                
\medskip
                                                                                      
    The apparent bolometric magnitude M(bol)   for KP2001-176, and 251 are determined by applying the bolometric correction BC( K ) to the m(K) reddening corrected magnitude (see  [ 18] for details):
    \medskip
    
    $   mbol. = m( K ) + BC( K ) $~~~~~~~~~~~~~~  ~~~~~~~~~ ~~~~~~~~~ ~~~~~~~ ~~~~~~~~~   ( 6 ) 

\medskip

In equation ( 6 ) we adopted the value BC(K) = 2.8 obtained using the calibration BC(K) vs. ( K-[12] ) index for O Ð rich Miras given in Fig. A3 of the same paper [18]. Note that IRAS 12 $\mu$m magnitude was calculated from the densities quoted in the catalogue following the prescription given in [19].   

    The results for KP2001-18, 176 and 251 are presented in Table 4.

 \begin{table} [h]
 \centering
\caption{Period, absolute magnitude and distance of  KP2001-18, 176, 251
}
\begin{tabular}{cccccccc}
\hline
\noalign{\smallskip}
KP2001 & Var.type& Period& m$_{bol}$& M$_{bol}$&M$_K$  &D(pc)       &D(pc)    \\
Number&               & Days  &mag.& mag.& mag&from M$_{bol}$& from M$_K$\\  
   18     &  SR         & 60      &  -     &  -     &-6.88$\pm$ 0.35& -  &810$\pm$100\\
   176   & Mira       & 330    & 8.71 &-4.71&-7.70$\pm$0.12&4850&5100$\pm$100\\
   251   &Mira        & 347   & 5.12 & -4.82&-7.82$\pm$0.09&980 &1100$\pm$150\\ 
\hline
\end{tabular}
\end{table}

For the stars, KP2001-16 and 243 the approximate distances can be estimated, using Guide Star Catalogue (GSC2.3) [15] (Vizier Online Catalogue ÐI/305) V magnitudes, and adopting absolute visual magnitudes for M2III=-1.1 (see for more details at http://www.handprint.com/ASTRO/ specclass.html\#luminositycodes) 

 Table 5 presents data for KP2001-16 and 243 stars
 
 \begin{table}
 \centering
\caption{Absolute magnitude and distance of KP2001-16 and 243
}
\begin{tabular}{cccc}
\hline
\noalign{\smallskip}
KP2001 & mv(GSC2.3)& M$_V$  & Distance \\
Number & mag.           & mag.      &pc            \\
16          &14.21           &-1.1        &2510        \\
243        &12.93           &-1.1        &2000         \\
\hline
\end{tabular}
\end{table}

\section{Concluding remarks.}   
In this paper, third in this series, we continue spectroscopic and photometric study oxygen-rich stars from KP2001 catalogue in Cepheus direction. To study our stars we followed the same methodological approach as in [1].  For seven new targets we obtained moderate resolution CCD spectra which allowed to determined spectral types and luminosity classes.  From the near infrared colors we measured the interstellar extinction; for the variable stars we could apply the Period-luminosity relations from which we then derived absolute magnitudes and distances.  All the variable stars studied in paper [1] and in the present paper have periods automatically determined by the NSVS survey, based on a relatively short monitoring. We already started a long-term monitoring not only to these stars, but to all the KP2001 stars in the Cepheus region in order to improve the value of the periods for the known variable and to look for variability in those not yet studied.  Actually we intend to apply the same analysis to all the 257 stars with the purpose of clarifying their nature and compare the ratio of Carbon to Oxygen-rich giant stars in this region of the Milky Way. 
In addition we are working on the infrared color Ð color diagrams involving IRAS, AKARI, WISE data, facing with the difficulties due to the interstellar extinction toward the Cepheus direction.

\medskip

{\bf Acknowledgments.} This work is based on observations obtained with the Cassini telescope of the Bologna observatory.

\medskip

REFERENCES

1. M.A.Kazaryan, G.V.Petrosyan, Astrofizika, 44, 413, 2001.

2. C.Rossi, S.Gaudenzi, G.V.Petrosyan et al., Astrofizika, 52, 577, 2009.

3. G.V.Petrosyan, Astrofizika, 56, 459, 2013.

4.  M.S.Bessell, J.M.Brett, Publ.Astron.Soc.Pacif., 100, 1134, 1988.

5. R.M.Sharples, P.A.Whitelock, M.W.Feast, Mon.Notic.Roy.Astron.Soc., 272, 139, 1995.

6. K.L.Cruz, I.Neill Reid et al., Astron. J., 126, 2421, 2003.

7. I.Neill Reid, A.J.Burgasser, K.L.Cruz et al., Astron. J., 121, 1710, 2001.

8. F.J.Zickgraf, J. Krautter, S.Reffert et al., Astron. J., 433, 151Z, 2005.

9. J.M. Carpenter, Astron. J., 121, 2851, 2001.

10. P.R. Wozniak, W.T. Vestrand, C.W. Akerlof, Astron. J., 127, 2436, 2004.

11. S. Ferraz-Mello, Astron. J., 86, 619, 1981. 

12. P.R. Wozniak, S.J. Williams, W.T. Vestrand et al., Astron. J., 128, 2965, 2004.

13. Sh. Deguchi, T. Sakamoto, T. Hasegawa, Publ. Astron. Soc. Japan, 64, 4D, 2012.

14. G. R. Knapp, D. Pourbaix, I. Platais, A. Jorissen, Astron. Astrophys., 403, 993, 2003.

15. B. Lasker, M. C. Lattanzi, B. J. McLean et al., Astron. J., 136, 735, 2008.

16. Whitelock P, Marang F. and Feast M., Mon. Not. R. Astron. Soc, 319, 728, 2000.

17. P.A.Whitelock, M. W. Feast and F. van Leeuwen, Mon. Not. R. Astron. Soc., 386, 313, 2008. 

18. T. Le Bertre, M. Matsura, J. M. Winters et al., Astron. Astrophys., 376, 997, 2001.

19. H. J. Walker, M. Cohen, Astron. J., 95, 1801, 1988.

\end {document}